\documentclass[prb,twocolumn,amsmath,amssymb,showpacs]{revtex4}
\usepackage{graphicx}

\def\beq{\begin{eqnarray}}
\def\eeq{\end{eqnarray}}
\def\beqa{\begin{eqnarray}}
\def\eeqa{\end{eqnarray}}

\begin{document}

\title{Cooperative effect of phonons and electronic correlations 
for superconductivity in cobaltates 
} 

\author{A. Foussats\dag, A. Greco\dag, M. Bejas\dag{}  and A. 
Muramatsu\ddag}
\affiliation{
{\dag} Facultad de Ciencias Exactas, Ingenier\'{\i}a y Agrimensura and 
Instituto de F\'{\i}sica Rosario
(UNR-CONICET). 
Av. Pellegrini 250-2000 Rosario-Argentina.\\
{\ddag} Institut f\"ur Theoretische Physik III, Universit\"at Stuttgart, 
Pfaffenwaldring 57, D-70550 Stuttgart, Germany.  
}

\date{\today}

\begin{abstract}
 
We propose that unconventional superconductivity in 
hydrated sodium cobaltate $Na_xCoO_2$ results from an interplay
of electronic correlations and electron-phonon interactions.
On the basis of the $t-V$ model plus phonons
we found evidences for a) unconventional superconductivity, b) 
realistic values 
of $T_c$ and c) the dome shape existing near $x \sim 0.35$.
This picture is obtained for  
$V$ close to the critical Coulomb repulsion $V_c$ which separates the
uniform Fermi liquid from $\sqrt{3} \times \sqrt{3}$ CDW ordered phase. 

\end{abstract}

\pacs{74.20.-z,74.20.Rp}

\maketitle

The discovery of superconductivity in 
$Na_xCoO_2.yH_2O$ at $T_c\sim 5K$ for $x=0.35$ and $y=1.3$  
\cite{Takada} has
generated a great interest in the solid state physics community.
Superconductivity is observed near doping 
$x \sim 0.35$ 
where $T_c$ follows a characteristic dome shape\cite{Ong}
(see also Ref.[\onlinecite{sakurai}]).

Cobaltates may be considered as electron-doped Mott insulators with 
layered structure where the $Co$ atoms are in 
a triangular lattice. Hence,
cobaltates  
are strongly correlated systems and  
Hubbard or 
$t-J$ models were derived 
to study these materials\cite{Baskaran}. 
The importance of strong electronic correlation is confirmed
by recent photoemission studies \cite{ARPES} showing for $x = 0.3$
a reduction of bandwidth 
by a factor of two with respect to the calculated ones \cite{LDA}

Several theoretical works predict  
unconventional superconductivity in cobaltates due to strong electronic 
correlation\cite{Baskaran,Lee1,Kumar,Tanaka,Kuroki,Lee}. 
Recent nuclear quadrupolar resonance (NQR) \cite{NQR}, 
$\mu$SR \cite{Higemoto} and specific heat \cite{SHeat} measurements
indicate that superconductivity is unconventional, with  
an order parameter in the spin-triplet channel, exclude time-reversal 
symmetry breaking, and are consistent with nodal lines,
such that the recently proposed next-nearest neighbor (NNN) 
$f$-wave triplet state \cite{Tanaka,Kuroki,Lee} seems a good
candidate. However, estimates of $T_c$ lead to extremely small values
\cite{Lee}. A natural extension to the proposed charge fluctuation
mechanism \cite{Tanaka,Lee} is to consider electron-phonon (e-ph) interaction,
since charge fluctuations could be enhanced by it. However, a conventional 
e-ph mechanism would be at odds with unconventional superconductivity.

We show below that the interplay between 
electronic correlations and the e-ph interaction 
close to a charge-density instability
lead to i) unconventional pairing, as originally proposed \cite{Tanaka,Lee}
but with ii) a realistic value for $T_c$ and iii) a dome shape around doping 
$x \sim 0.35$. 

We start with the $t-V$ model
plus phonons\cite{Jcero}:
\begin{eqnarray}
H & = & - t \sum_{<ij>,\sigma}\;( {\tilde{c}\dag}_{i\sigma} 
\tilde{c}_{j \sigma}
+ h.c.) +
V \sum_{<ij>} n_i n_j
\nonumber \\ & & +\sum_i \omega_0 (a\dag_{i} a_i+\frac{1}{2})
+g \sum_i (a\dag_i+a_i)n_i
\end{eqnarray}
\noindent
where $t$ and $V$ are the hopping and Coulomb repulsion 
between the nearest-neighbors (nn) sites $i$ and $j$ on the triangular 
lattice.   
${\tilde{c}\dag}_{i \sigma}$ and $\tilde{c}_{i \sigma}$ 
are the fermionic creation and 
destruction operators for holes, respectively, under the constraint 
that double occupancy is excluded. $n_i$ is the fermionic density. 
$a\dag_{i}$ and $a_i$ are the phonon
creation and destruction operators, respectively.  
The e-ph coupling $g$ as well as the phonon frequency $\omega_0$ 
are considered to be constants. The interaction $V$ is motivated by 
optical absortion experiments \cite{optical}
indicating charge instabilities at $x=0.25$
and $x=0.5$, i.e.\ around the region where superconductivity takes place.

To study the model (1) we will use a recently developed path integral 
large-N approach for Hubbard operators \cite{Adriana}, 
starting with
the well known 
relations \cite{Hubb} between the $\tilde{c}$'s and 
the Hubbard operators $X$; 
$X^{\sigma 0}_i={\tilde{c}\dag}_{i \sigma}$
and 
$n_i=(
X^{\uparrow \uparrow}_i+
X^{\downarrow \downarrow}_i)$. 

Our large-N path integral approach for Hubbard operators 
does not require any decoupling scheme as in slave bosons, 
and therefore, problems with 
gauge fluctuations or Bose condensation are avoided.
At leading order, our formalism is equivalent to the slave-boson 
approach. However, at the next to leading order (which is necessary 
to calculate dynamical properties) the 
two formulations differ.
Our formalism was recently tested by comparing 
dynamical properties with exact 
diagonalization on small clusters \cite{Jaime}.  

In
order to get a finite theory in the $N\rightarrow \infty$ limit,
we rescale $t$ to $t/N$, 
$V$ to
$V/N$ and $g$ to $g/\sqrt{N}$ and the spin projections $\sigma$ 
are extended from $2$ to $N$.   
As in slave-boson, to leading order, we have free fermions with 
an electronic band, 
$E_{k}=  
-tx(\cos k_x+2 \cos \frac{\sqrt{3}}{2}k_y \cos \frac{1}{2}k_x)$, 
renormalized by Coulomb interactions 
where $x$ is the electron doping.   
In the following $t$ is considered to be $1$. From this electronic
dispersion we obtain a large Fermi surface (FS) enclosing the $\Gamma$
point.  
First principles calculations 
\cite{LDA}  predict, apart 
from this FS, the existence of small pockets near the $K$ points. 
This last picture is of interest for some 
studies were the pockets are relevant for superconductivity \cite{pokets}.
However, it is important to notice that recent ARPES experiments \cite{ARPES}
do not show the presence of pockets.
A theoretical explanation for the absence of pockets was given using 
$LDA+U$ \cite{LDAU}. 

\begin{figure}
\begin{center}
\setlength{\unitlength}{1cm}
\includegraphics[width=6cm,angle=0.]{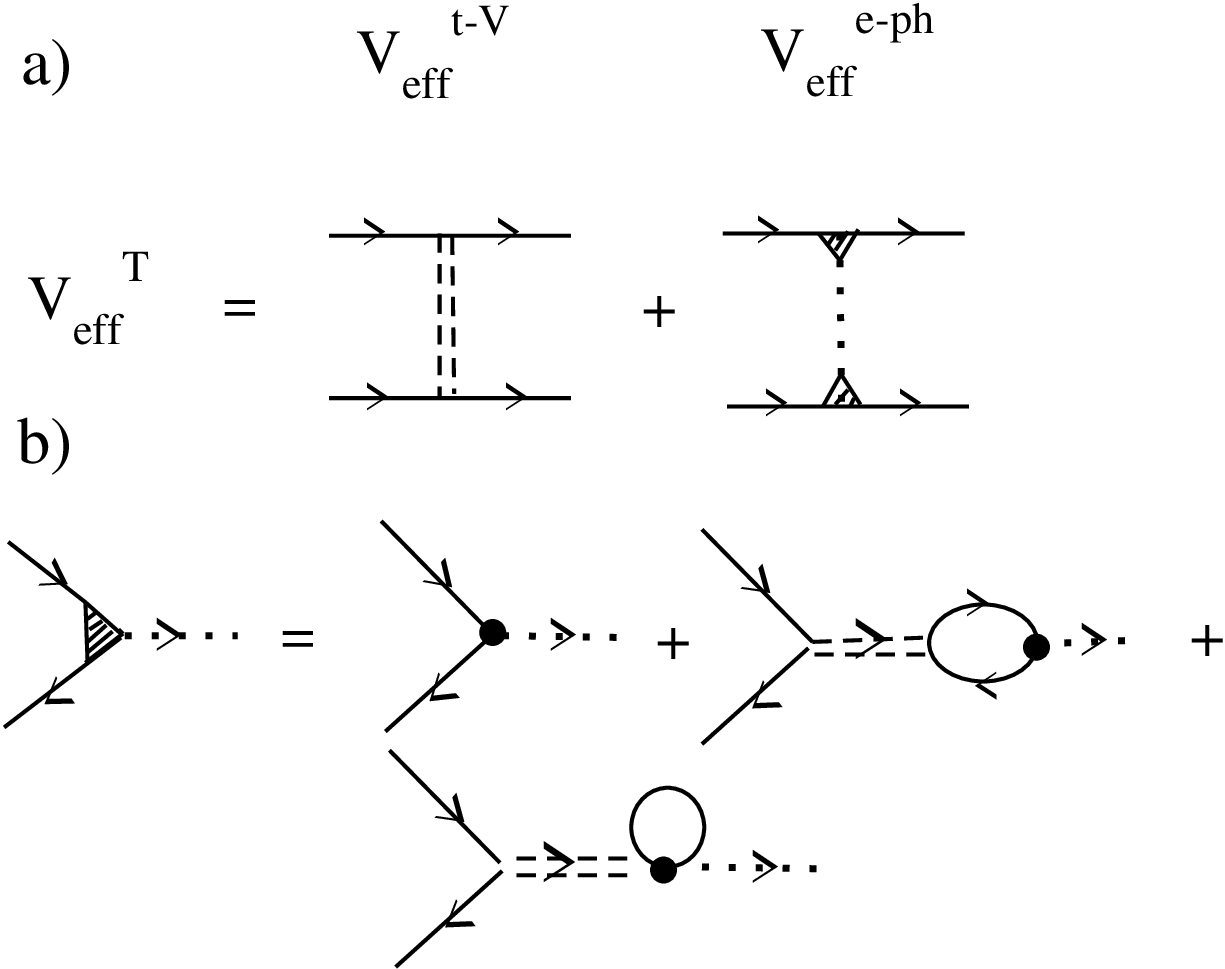}
\end{center}
\caption{a) Total effective paring $V_{eff}^T$ 
as the sum of a pure electronic 
mediated $V_{eff}^{t-V}$ and a phonon mediated $V_{eff}^{e-ph}$ 
interactions. Solid, double dashed, and dotted lines are the propagators 
for fermions, charge fluctuations in the $t-V$ model 
and phonons, respectively.
In $V_{eff}^{e-ph}$, the bare e-ph vertex, 
g (solid circle), is renormalized by 
electronic correlations as showed in b).  
The last diagram contains a four leg vertex 
proportional to $g$ which is generated when our X-operator approach 
is applied to the last term of the Hamiltonian (1). The Feynman rules 
used in the evaluation of the diagrams are those presented in 
Refs.[\onlinecite{Adriana, Jaime}] and they are given in terms of 
X-operators and not in terms of slave particles. 
 }
\end{figure}

Using our Feynman rules \cite{Adriana,Jaime} 
we can calculate the pairing diagrams of Fig.\ 1a. The total pairing 
effective interaction $V_{eff}^T(\vec{k}-\vec{k'})$ 
is the sum of a pure electronic  
$V_{eff}^{t-V}(\vec{k}-\vec{k'})$ 
and phonon 
$V_{eff}^{e-ph}(\vec{k}-\vec{k'})$ 
terms .
The charge fluctuation propagator (double dashed line) in $V_{eff}^T$
contains a RPA-type series of electronic bubbles, such that 
$V_{eff}^T$ is exact in $O(1/N)$\cite{Adriana,Jaime}.

In Ref.[\onlinecite{Jaime}], 
it was shown that the pairing effective interaction from 
the pure $t-V$ model is of $O(1/N)$.
Due to the rescaling of the e-ph interaction, $g$, 
superconductivity from phonons also appears at $O(1/N)$ and therefore 
it can be treated on an equal footing to superconductivity in 
the pure $t-V$ model.

Pairing is mediated mainly by charge fluctuations \cite{Jaime}
(double dashed line) in $V_{eff}^{t-V}$ (first diagram of Fig.\ 1a).
As will be seen later, for the parameters used, the renormalized e-ph 
coupling (Fig.\ 3) is still in the weak coupling regime, so that
corrections to the self-energy of the phonons can be disregarded as usual.
The new contribution to the e-ph pairing potential is the 
vertex (dark triangle) in the second diagram of Fig.\ 1a which 
represents the e-ph interaction renormalized by electronic correlations.
The diagrams of Fig.\ 1b show that the renormalization
of the bare vertex $g$ is due to the electronic correlations of the 
pure $t-V$ model, that will give the main contribution to our results.  

For a given doping, $x$, the leading order uniform 
Fermi liquid is 
unstable against a $\sqrt{3} \times \sqrt{3}$ CDW for $V$ 
larger than a critical value 
$V_c$. 
Except for a factor $2$, (due to different 
definition of the Coulomb term in the Hamiltonian) 
our mean-field phase diagram in the $V_c-x$ plane agrees with the 
obtained one in Ref.\ [\onlinecite{Lee}] (see Fig.\ 2 of that paper).
For instance, in our case, $V_c \sim 1.1$  
for the doping $x=1/3$. 
However, for $x \sim 1/3$, the instability cannot be seen as a softening
of a collective charge mode, but a redistribution of spectral weight
takes place. For the values of V considered here, still a clearly 
defined mode appears at high energy ($\sim 3t$) around the wavevector 
$(4\pi/3,0)$ but sizeable spectral weight appears on a broad structure at 
low energies with a maximum at a scale $\sim t/2$ (see inset (b) in Fig.\ 2). 
The inclusion of an antiferromagnetic exchange interaction $J$ brings almost 
no quantitative changes, as expected for the rather hight doping level, and
consistent with a paramagnetic metal \cite{Ong1}.   
Further details of the 
density response close to the instability will be given 
elsewhere \cite{Ale}.
We remind that the $V_c-x$ line separating 
the uniform Fermi liquid from the $\sqrt{3} \times \sqrt{3}$ CDW 
phase has a parabola-like shape with a minimum closer to 
the doping $x \sim 0.35$ where superconductivity 
takes place in cobaltates. 
The proximity to a charge instability seems to be confirmed in
optical experiments \cite{optical}, where an anomalous behavior, such as 
the appearance of a low energy peak is observed. A similar situation 
appears in one quarter 
filling organic materials and has been interpreted as being induced by
charge fluctuations close 
to the charge-ordered transition \cite{Jaime1}.

\vspace{0.5cm}
\begin{figure}[h]
\begin{center}
\setlength{\unitlength}{1cm}
\includegraphics[width=8.0cm,angle=0.]{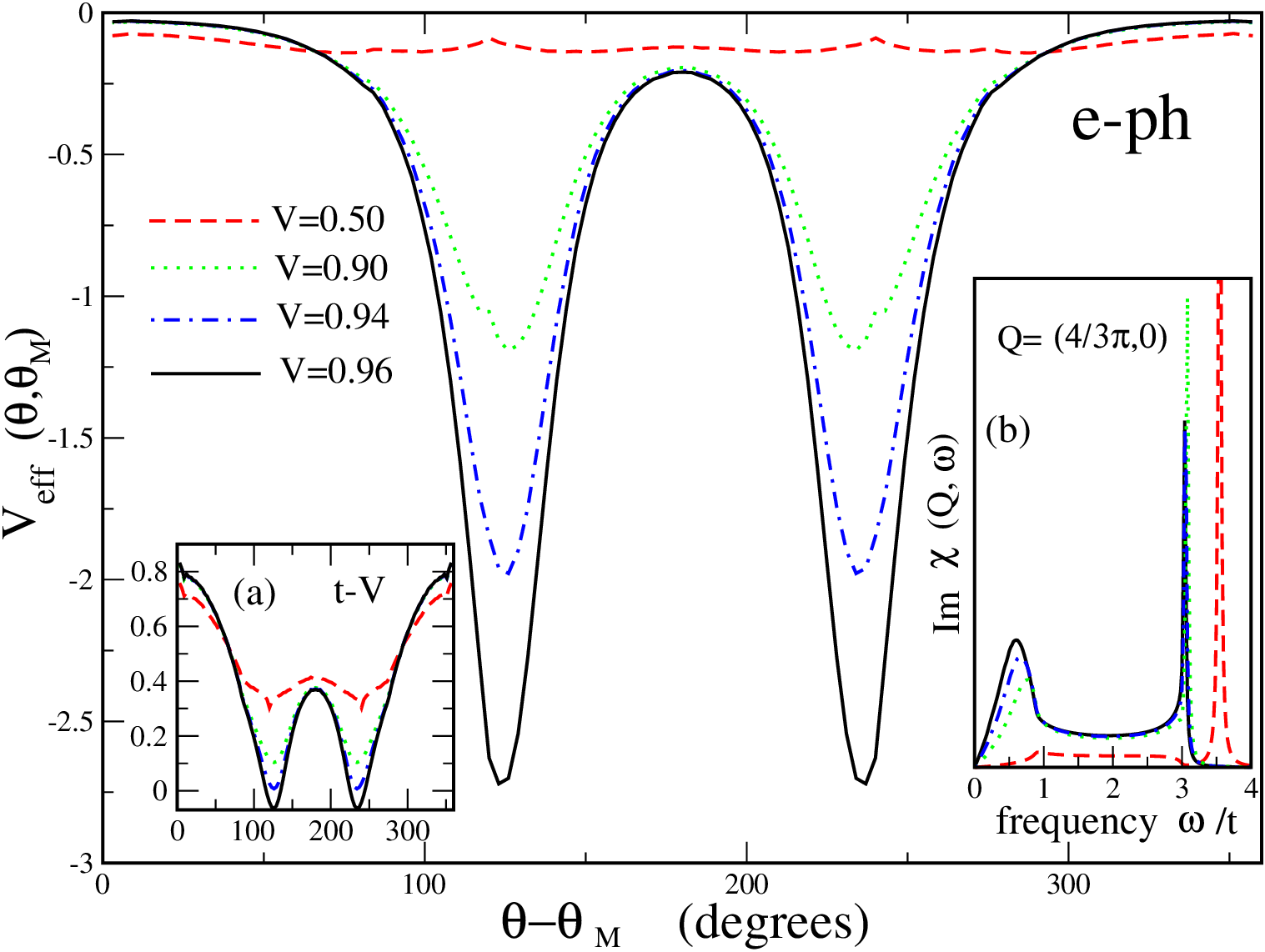}
\end{center}
\caption{
(color online) 
Behavior of 
$V_{eff}^{e-ph}(\vec{k}-\vec{k'})$ 
for different $V$'s approaching 
$V_c$. 
As in Ref.\ [\onlinecite{Lee}] $\vec{k'}$ is fixed on the FS 
at the angle $\theta_M=\pi/6$ and $\theta$ measures the angle of
$\vec{k}$ which is always on the FS. 
When $V$ approaches $V_c$, $V_{eff}^{e-ph}$ becomes more  
anisotropic and more attractive favoring 
triplet $NNN f$-wave superconductivity.
Inset (a):
behaviour of $V_{eff}^{t-V}(\vec{k}-\vec{k'})$. Inset (b): imaginary 
part of the density response
$\chi(\vec{Q},\omega)$, at $\vec{Q}=(4/3 \pi,0)$,  
for different $V$ (the same of the main panel) approaching 
$V_c$ showing the redistribution of the spectral weight discussed 
in the text. 
}
\end{figure}

In Fig.\ 2 we show $V_{eff}^{t-V}$ and 
$V_{eff}^{e-ph}$  
for two $\vec{k}$-vectors ($\vec{k}$ and $\vec{k'}$)
on the Fermi surface for $x=1/3$ and for different 
Coulomb repulsion $V$, approaching $V_c$. As in Ref.\ [\onlinecite{Lee}],
the momentum $\vec{k'}$ is fixed at the angle $\theta_M=\pi/6$ and 
$\vec{k}$ 
runs around the 
Fermi surface. 
The effective interaction $V_{eff}^{t-V}$, 
showed in inset (a), agrees 
with that obtained 
in Fig.\ 3b of Ref.\ [\onlinecite{Lee}]. When $V$ approaches $V_c$, the 
effective interaction $V_{eff}^{t-V}$ 
becomes more anisotropic and more attractive,  
favoring superconductivity in the triplet $NNN f$-wave (see below) in
agreement with Ref.\ [\onlinecite{Lee}].

Figure 2 (main panel) shows our main result. 
The bare e-ph paring effective interaction 
which is originally isotropic 
(the bare $g$ and $\omega_0$ were considered 
to be constants)  
becomes strongly $\vec{k}$-dependent 
due 
to vertex corrections by electronic correlations.
In the calculation of $V_{eff}^{e-ph}$ we used the bare value 
$g$ obtained from $\lambda=2 g^2 N(0)/\omega_0$ where $N(0)$ is 
the bare fermionic density of states and, the dimensionless bare 
e-ph coupling 
$\lambda$ was chosen to be $0.4$ \cite{Fano}.
Vertex corrections obtained by us using the large-N path integral 
approach are very similar to the early calculation of 
Ref.\ [\onlinecite{Kulic}] which were 
used before for studying transport \cite{Zeyher} and isotope 
effect in cuprates \cite{Greco}.

The main effect of correlations occurs for $V$ close to the
critical value $V_c$ where the e-ph pairing interactions 
$V_{eff}^{e-ph}$ is strongly anisotropic following the same 
shape of $V_{eff}^{t-V}$ (see Fig.2).
Therefore, charge fluctuations close to the 
$\sqrt{3} \times \sqrt{3}$ CDW order enhance the anisotropy of 
$V_{eff}^{e-ph}$ with the same symmetry as $V_{eff}^{t-V}$ and 
reinforce it. 

We use the effective potentials to compute the dimensionless 
effective couplings in the different pairing channels or
irreducible representations of the order parameter. 
The critical temperature $T_c$ can be then estimated 
from $T_{ci}=1.13 \omega_c exp(1/\lambda_i)$, where $\omega_c$ 
is a suitable cutoff frequency, and $i$ denotes the different pairing channels.

The effective couplings $\lambda_i$  are defined as
\cite{Jaime}:

\begin{eqnarray}
\lambda_i= 
\frac{\int (d {\bf k} /|v_{\bf k}|) \int (d {\bf k'}/|v_{\bf k'}|)
g_i({\bf k'})
V_{eff}({\bf k'-k}) g_i({\bf k})}
{ 
(2 \pi)^2 
\int (d {\bf k}/|v_{\bf k}|)  g_i({\bf k})^2 }
\end{eqnarray}

\noindent
where the functions $g_i({\bf k})$, encode the different pairing
symmetries (see Table I of Ref.[\onlinecite{Lee}] for the 
triangular lattice), and $v_{\bf k}$ are the quasiparticle velocities at
the Fermi surface. The integrations are restricted to the Fermi
surface. $\lambda_i$ measures the strength of the interaction
between electrons at the Fermi surface in a given symmetry channel
$i$. If $\lambda_i > 0$, electrons are repelled. Hence,
superconductivity is only possible when $\lambda_i <0$.

\begin{figure}[h]
\begin{center}
\setlength{\unitlength}{1cm}
\includegraphics[width=7cm,angle=0.]{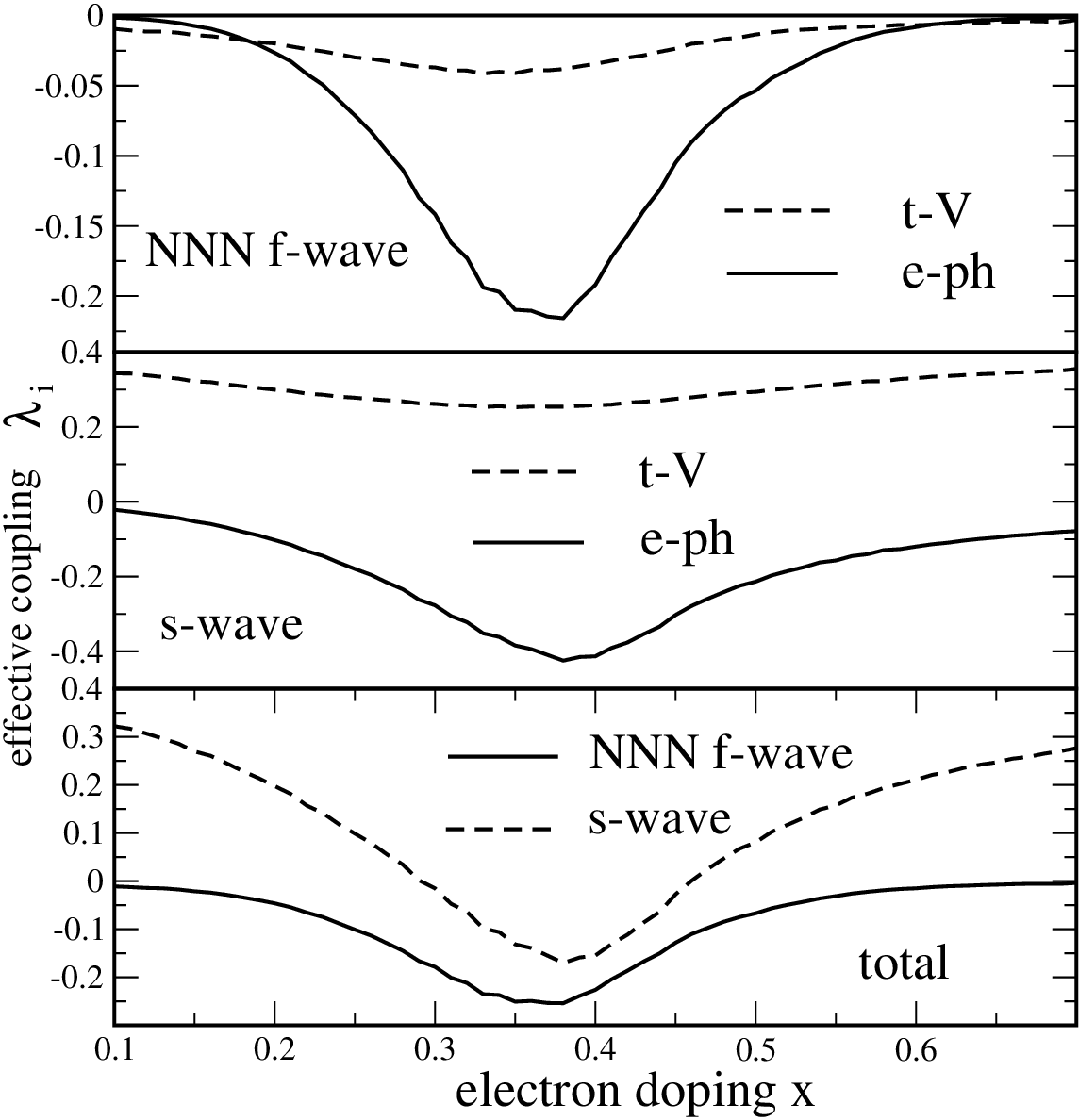}
\end{center}
\caption{
Dimensionless superconducting coupling $\lambda_i$
for the pure electronic and phononic models as well as 
the total case. $\lambda_i$ is calculated for the 
$s$-wave and $NNN f$-wave symmetries and for $V=0.96$ and $\lambda=0.4$. 
The other
channels (not shown) are not relevant.  
 }
\end{figure}

In Fig.3 we show $\lambda_i$ as a function of 
doping for the pure $t-V$-mediated, the renomalized e-ph mediated, and total
($\lambda_i^T=\lambda_i^{t-V}+\lambda_i^{e-ph}$) 
effective interactions in the most relevant channels $s$-wave and 
$NNN f$-wave for $V=0.96$ and $\lambda=0.4$. 
For pure electronic pairing, in agreement with 
Ref.[\onlinecite{Lee}], the lowest effective coupling has triplet 
$NNN f$-wave symmetry near $V_c$ (dashed line in the upper panel). 
As shown by the dashed line in the upper panel of Fig.\ 3, 
the lowest value of $\lambda_i \sim -0.05$ leads to 
$exp(-1/0.05) \sim 10^{-9}$. This extremely small value 
means that a value for $T_c$ of the order of a few Kelvin 
is only possible for an unrealistic high value for the cutoff 
$\omega_c$.
The situation does not change very much in the most favorable 
case when the system is very close to the $\sqrt{3} \times \sqrt{3}$
CDW order (or $V=1 \sim V_c$). Even in this case the value of $\lambda_i
\sim -0.1$ is also very unfavorable for superconductivity.
Experimentally, the dome in $T_c$ around $x \sim 0.35$ has a finite 
extent and it is not limited to a single value of doping, signaling that 
the range of interactions $V$ able 
to lead to superconductivity cannot be only limited to 
values very close to $V_c$.  

The situation is different for the phonon-mediated case.
The solid line in the upper panel of Fig.\ 3 shows $\lambda_i$ 
in the $NNN f$-wave.
The absolute values of $\lambda_i$ 
are clearly larger than those from the pure electronic case.
For $x \sim 0.36$, $\lambda_i \sim -0.22$.
Furthermore, the phonon-mediated case leads to a more pronounced 
dome shape in the doping range around $x \sim 0.35$. 

In the middle panel we present similar results to those of the 
upper panel 
for the $s$-wave symmetry.
The solid line shows the results for the phonon-mediated 
case with large $\lambda_i \sim -0.4$.
However, the results for the $t-V$ model (dashed line) 
are strongly repulsive, $\lambda_i \sim 0.3$, diminishing the 
attractive effect given by phonons.

The lowest panel shows results for $\lambda_i^T$.  
The 
lowest effective coupling is in $NNN f$-wave symmetry following 
a dome shape around $x \sim 0.35$ where  
$\lambda_i^T$ takes values $\lambda_i^T \sim -0.25$ which are larger
($\sim$ six  times) 
than the corresponding ones from the pure electronic model.
This means that $T_c$ can be of the order of a few Kelvin 
for a cutoff frequency of the order of a high phonon mode.
The Debye frequency in cobaltates 
is expected to be large according to recent first principles 
lattice dynamics calculations \cite{Li} that shows the 
existence of optical 
phonons as high as 
$75 meV$. Certainly it is no our aim to give a quantitative 
value of $T_c$ but to show that the present treatment leads to realistic 
scales in critical temperature and doping.

At this point one remark is in order. With increasing $\lambda$, 
the total superconducting couplings $\lambda_i^T$ in the $s$-wave 
and $NNN f$-wave channels become more attractive and, for $\lambda > 1$
both symmetries are nearly degenerated. However, lacking detailed information
about the e-ph coupling, we take a cautious value for $\lambda$, that is
already sufficient to trigger superconductivity, and as shown above, of
unconventional type.     

As superconductivity has a large contribution from phonons we 
expect a large isotope coefficient. 
Our theory also predicts a rather constant isotope coefficient 
along the dome in contrast to the strong doping 
dependent isotope coefficient 
in cuprates \cite{Frank}. 
To our knowledge, isotope effect experiments 
are still not available for cobaltates.

It was recently proposed \cite{Kotliar}, 
and studied ex\-perimen\-tally \cite{sakurai}, that hydration causes the 
electronic structure to be more two dimensional. 
We think that due to this effect the 
Coulomb repulsion $V$ is less screened when the system is hydrated. 
If $V$ is small when the system is not hydrated, phonons will favor 
superconductivity only in the $s$-wave channel. However, the strong
repulsion 
with this symmetry from the $t-V$ mediated interaction 
will cancel the pairing from phonons. This may be the 
reason for the nonexistence of superconductivity in nonhydrated 
cobaltate. 
Note that in Ref.\ [\onlinecite{sakurai}] it was shown that $T_c$ decreases 
with decreasing lattice parameter $c$. Finally we would like to stress
that the phonon contribution to superconductivity, differently to the pure 
$t-V$ case, does not need the system to be very 
close to the charge ordered 
phase. In fact, we find paring inside the metallic phase, in fully
agreement with the fact that charge order was never observed for doping
$x \sim 0.35$
  
In summary, we have shown that  
a cooperative effect of 
phonons and electronic correlations may lead to unconventional 
superconductivity in cobaltates. 
We showed that superconductivity is possible 
due to e-ph interaction where the bare e-ph
coupling is renormalized by electronic correlations in the proximity 
of the $\sqrt{3} \times \sqrt{3}$ CDW instability of the uniform 
Fermi liquid. 
Furthermore, we have shown  
that superconductivity is possible for a realistic value 
of $T_c$, and we also obtained the characteristic dome shape near 
$x \sim 0.35$.


\begin{thebibliography}{99}
\bibitem{Jcero}   
We have done similar calculations 
for the case of finite $J$. 
However, for $J/t \sim 0.1$ (i.e.\ for experimentally
relevant values \cite{ARPES}), the situation discussed here holds. 
(We will present more details in a future publication).
\end{thebibliography}
\end{document}